\documentclass[a4paper,fleqn,usenatbib]{mnras}

\usepackage[T1]{fontenc}
\usepackage{ae,aecompl}

\usepackage{graphicx}
\usepackage{amsmath}
\usepackage{amssymb}

\usepackage{color}
\usepackage{ulem}

\title[Formation of cluster's central compact dwarfs]{Two channels for the formation of compact dwarf galaxies in clusters of galaxies}

\author[N. Martinovi\'c and M. Micic]{
Nemanja Martinovi\'c,$^{1}$\thanks{E-mail:nmartinovic@aob.rs}
Miroslav Micic$^{1}$
\\
$^{1}$Astronomical Observatory, Volgina 7, 11060 Belgrade, Serbia
}

\date{Accepted XXX. Received YYY; in original form ZZZ}

\pubyear{2017}

\begin{document}
\label{firstpage}
\pagerange{\pageref{firstpage}--\pageref{lastpage}}
\maketitle

\begin{abstract}
We have identified two channels for the formation of compact dwarf galaxies in the Illustris simulation by reconstructing mass and distance histories of candidates located in the vicinity of the simulation's most massive cluster galaxies. One 
channel is tidal stripping of Milky Way mass galaxies that form outside of clusters and eventually sink into them, spiraling in toward central massive objects. Second channel of formation is an in-situ 
formation (in reference to the parent cluster) of dwarf mass galaxies, with negligible evolution and limited change in stellar mass. We find 19 compact dwarf galaxies at the centers of 14 clusters, consistent with observations. 30\% of them have external origin while 70\% are formed in-situ.
\end{abstract}

\begin{keywords}
galaxies: dwarf --
galaxies: clusters: general --
galaxies: formation --
galaxies: evolution --
galaxies: interactions --
galaxies: elliptical and lenticular, cD
\end{keywords}



\section{Introduction}

In recent years, increased interest in dwarf galaxies lead to discovery of new classes of compact dwarfs, populating the link between globular clusters (GCs) and dwarf ellipticals (dEs) in the galaxy mass-size diagram. 

Since 1999 galaxies classified as ultra-compact dwarfs (UCDs) were identified in Fornax \citep{Hilker1999, Drinkwater2000, Drinkwater2003} and subsequently discovered in several other clusters as well: Abell 1698 \citep{Mieske2004}, Virgo \citep{Jones2006}, Coma \citep{Price2009}, Hydra I \citep{Misgeld2011}, Perseus cluster \citep{Penny2012, Penny2014}. 

Apart from UCDs, in recent years additional M32-like compact ellipticals (cEs) have been found as well \citep{Mieske2005, Chilingarian2007, chilingarian2008, Zhang2017}, additionally bridging the gap between GCs, UCDs and dEs. Most of M32-like cEs were identified in the vicinity of the most massive galaxies inside the clusters. 

Abundance of findings of new compact dwarf galaxies lead to the question of their formation. Considering the heterogeneity of these objects it is expectable that their formation histories will differ significantly.
Indeed, several channels for their formation were proposed - for UCDs: that they represent overmassive GCs \citep{Mieske2002}, that they form through mergers of massive stellar clusters \citep{Fellhauer2002},  tidal stripping of dwarf
galaxies \citep{Bekki2003}, while for compact dwarfs debated formations were: formation through tidal stripping of spiral galaxies \citep{Bekki2001} or simply in-situ formed objects whose growth is obstructed due to cluster membership \citep{Wellons2016}.

In this paper we reconstruct the formation history of compact dwarf candidates present at $z=0$ in the vicinity of most massive members of clusters of galaxies in Illustris-1 simulation as a mean of identifying channels of their formation. We have identified two channels: one that gives rise to the population which has externally (in reference to the final cluster) created progenitor galaxies; and the channel that gives rise to the population which has in-situ created progenitors.

\begin{table*}
	\centering
	\caption{Parameters for compact dwarf galaxy candidates from Illustris-1 simulation at $z=0$: label on plot (if applicable), distance to the most massive cluster galaxy ($kpc$), total mass (given in log), total stellar mass (given in log), maximum stellar mass of the candidate during its existance (given in log), maximum total mass of the candidate during its existance (given in log), stellar half-mass radius ($pc$), a 2D projected stellar half-mass radius ($pc$) (averaged over 1000 projections), minimal retrieved 2D projected stellar half-mass radius ($pc$) and maximum retrieved 2D projected stellar half-mass radius ($pc$). Above the horizontal line are parameters for external population and below the line for the in-situ population. }
	\label{tab:table}
	\begin{tabular}{cccccccccccc}
		\hline
		Label & d & & $\log(M_\odot)$ & $\log(M_\odot)$ & $\log(M_\odot)$ & $\log(M_\odot)$ & & $R$ & $R_{proj}$ & $R_{min}$ & $R_{max}$ \\
		& & & ($z=0$ total) & ($z=0$ stellar) & (max stellar) & (max total) & &  & & & \\

		\hline
& 45  &  & 9.3 & 9.3 & 10.3 & 12.0 &  & 1136 & 857  & 778  & 928  \\
& 91  &  & 9.0 & 9.0 & 10.5 & 11.9 &  & 1074 & 822  & 757  & 892  \\
& 99  &  & 9.4 & 9.3 & 10.8 & 12.1 &  & 1395 & 1049 & 1016 & 1078 \\
& 21  &  & 8.1 & 8.1 & 10.0 & 11.4 &  & 656  & 477  & 419  & 527  \\
& 92  &  & 9.8 & 9.8 & 10.8 & 12.0 &  & 1257 & 938  & 888  & 994  \\
& 73  &  & 8.8 & 8.8 & 9.9  & 11.5 &  & 943  & 731  & 626  & 855  \\ \hline
a) & 71  &  & 9.7 & 9.7 & 9.8  & 10.3 &  & 739  & 565  & 374  & 685  \\
b) & 32  &  & 9.4 & 9.4 & 9.7  & 10.0 &  & 558  & 427  & 338  & 479  \\
c) & 76  &  & 9.3 & 9.3 & 9.6  & 10.0 &  & 692  & 540  & 332  & 636  \\
d) & 71  &  & 8.8 & 8.8 & 9.4  & 9.9  &  & 587  & 449  & 432  & 463  \\
e) & 83  &  & 8.9 & 8.9 & 9.6  & 10.2 &  & 679  & 522  & 411  & 615  \\
f) & 80  &  & 8.5 & 8.5 & 9.0  & 9.6  &  & 895  & 699  & 567  & 778  \\
g) & 64  &  & 9.0 & 9.0 & 9.4  & 9.9  &  & 730  & 555  & 479  & 634  \\
h) & 85  &  & 8.6 & 8.6 & 9.2  & 9.7  &  & 808  & 614  & 538  & 688  \\
i) & 76  &  & 9.0 & 8.9 & 9.3  & 9.7  &  & 618  & 471  & 452  & 493  \\
j) & 104 &  & 8.5 & 8.5 & 8.9  & 9.7  &  & 1125 & 873  & 813  & 926  \\
k) & 88  &  & 8.5 & 8.5 & 8.9  & 10.0 &  & 847  & 655  & 563  & 741  \\
l) & 93  &  & 8.5 & 8.5 & 9.3  & 9.8  &  & 787  & 610  & 516  & 669  \\
m) & 58  &  & 8.7 & 8.7 & 9.1  & 9.8  &  & 967  & 739  & 548  & 855  \\
		\hline
	\end{tabular}
\end{table*}

\begin{figure}
	\includegraphics[width=1.0\columnwidth]{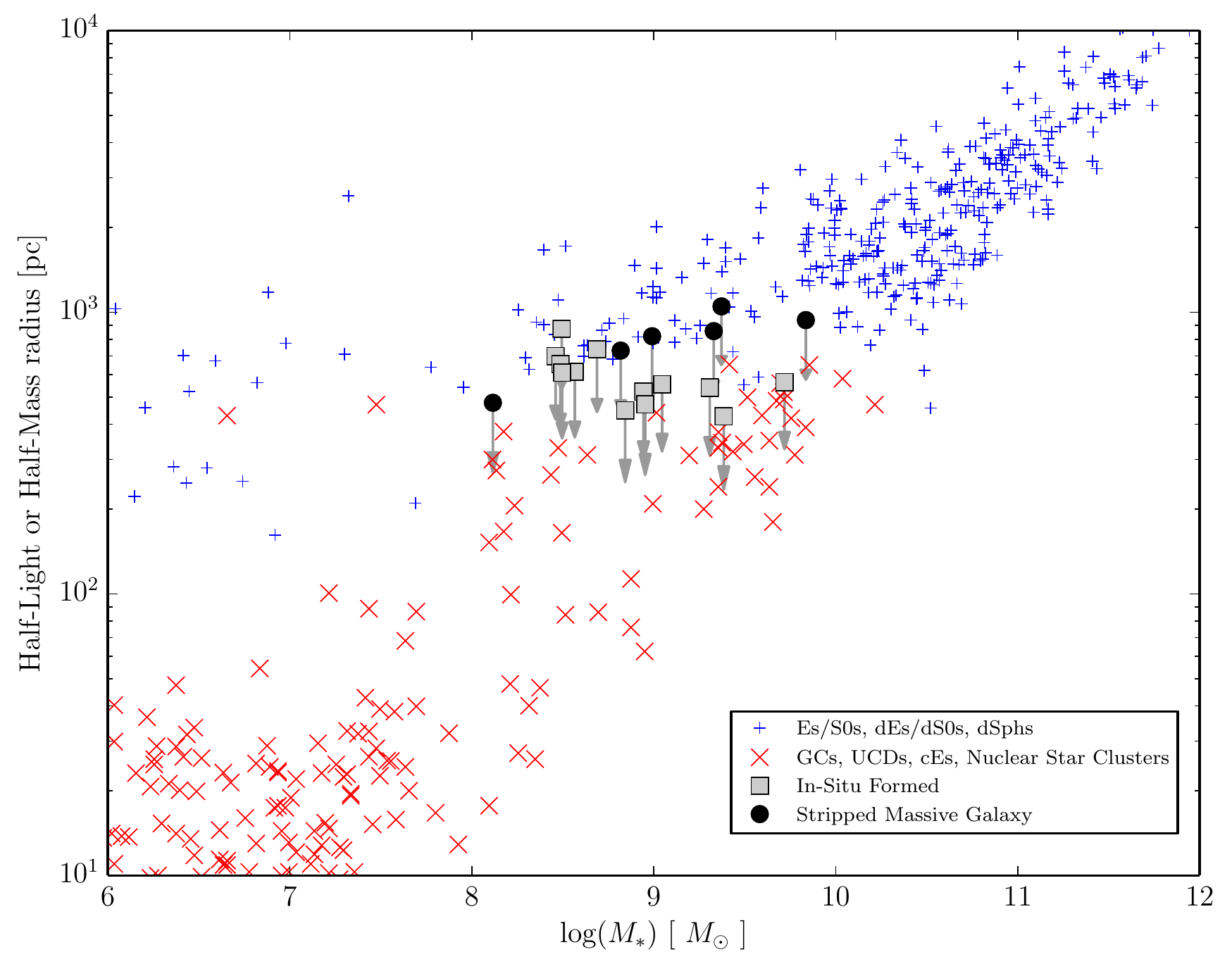}
    \caption{Galaxy mass-size diagram for compact dwarf galaxies. Blue pluses and red crosses are data points from \citet{Norris2014} presented with half-light radius. Overplotted are candidate compact dwarfs at $z=0$ from Illustris-1 presented with 2D projected half-mass radius (averaged over 1000 projections). Black circles represent objects which are remnants of tidally stripped Milky Way mass galaxies, while gray squares represent remnants of dwarf-like objects formed inside clusters of galaxies. Arrows convey that for these populations half-light radius is smaller than half-mass radius.}
    \label{fig:figure1}
\end{figure}

\section{Simulation}

For this work we use the results from the Illustris-1 simulation, a cosmological hydrodynamical simulation with a complete model for galaxy formation physics. It includes the formation of  stars with galactic super-winds driven by star-formation as an effects on their environments,  stellar evolution with chemical enrichment and stellar mass loss, as well as feedback from AGNs and formation and evolution of SMBHs. It postulates a $\Lambda CDM$ cosmology with the following parameters:
$\Omega_{m} = 0.2726$, 
$\Omega_{\Lambda} = 0.7274$,
$\Omega_{b} = 0.0456$,
$\sigma_{8} = 0.809$,
$n_{s} = 0.963$,
and $h = 0.704$. 
Initial conditions are generated at z = 127 in a periodic box with a side length of 106.5 Mpc, with initial gas temperature set to 245 K \citep{Vogelsberger2014}. 

Very high number of baryonic and dark matter elements of $2 \times 1820^{3}$ gives outstanding baryonic mass resolution of $1.26 \times 10^{6} M_{\odot}$ and dark matter mass resolution of $6.26 \times 10^{6} M_{\odot} $. Gravitational softening for Illustris-1 for baryons at $z=0$ is 710 pc, and the smallest hydrodynamical gas cell resolution is 48 pc \citep{Vogelsberger2014}. All the compact galaxy candidates in this paper are resolved with at least 300 particles at $z=0$.

Halo identification in the simulation was performed using the FOF algorithm with a standard linking length of 0.2 times the mean particle
separation, where 32 particles were used as the minimum number of particles for a halo. Subsequent refinement was done with the SUBFIND algorithm \citep{Springel2001} used to identify the gravitationally bound substructures. Merger tree was created using the SUBLINK code \citep{Rodriguez2015}. Results from the simulation have been made available through public release of the data \citep{Nelson2015}.

\section{Method}


\begin{table*}
\centering
\caption{Parameters for compact dwarf galaxy candidates from Illustris-1 simulation at their earliest progenitor formation time (given in second column): label on plots, first detection redshift, mass of gas (given in log), mass of stars (given in log), total star formation rate, fraction of HI, metallicity, metallicity given in solar units, maximum temperature for that cloud of gas, minimum temperature for that cloud of gas, average temperature for that cloud of gas.}
\label{tab:table2}
\begin{tabular}{ccccccccccc}
\hline
Label & z & $\log(M_\odot)$ & $\log(M_\odot)$ & SFR & $HI (\%)$ & $Z$ & $Z/Z_{0}$ & $T_{max}$ & $T_{min}$ & $\bar{T}$ \\
& & Gas & Stars & [$M_{\odot} /\ year$] & & & & [$10^{5} K$] & [$10^{5} K$] & [$10^{5} K$]\\
\hline
a)    & 0.73 & 10.3 & 9.0   & 29  & 0.10   & 0.019 & 1.5    & 4          & 0.01       & 0.8            \\
b)    & 0.62 & 10.0 & 8.9   & 10  & 0.03   & 0.025 & 2.0    & 2          & 0.03       & 0.7            \\
c)    & 0.7  & 9.9  & 9.3   & 21  & 0.02   & 0.026 & 2.0    & 4          & 0.02       & 1.3            \\
d)    & 0.95 & 9.6  & 8.6   & 4   & 0.10   & 0.022 & 1.7    & 2          & 0.01       & 0.6            \\
e)    & 1.67 & 10.2 & 9.0   & 12  & 0.08   & 0.011 & 0.8    & 3          & 0.02       & 0.5            \\
f)    & 1.41 & 9.5  & 8.1   & 1   & 0.33   & 0.007 & 0.5    & 1          & 0.04       & 0.3            \\
g)    & 1.9  & 9.8  & 9.2   & 23  & 0.00   & 0.015 & 1.2    & 4          & 0.02       & 1.7            \\
h)    & 2.58 & 9.5  & 9.1   & 10  & 0.01   & 0.020 & 1.6    & 4          & 0.06       & 1.5            \\
i)    & 1.6  & 9.5  & 9.1   & 10  & 0.02   & 0.027 & 2.1    & 3          & 0.06       & 1.4            \\
j)    & 0.85 & 9.3  & 7.9   & 1   & 0.27   & 0.015 & 1.2    & 1          & 0.02       & 0.3            \\
k)    & 1.9  & 9.8  & 8.1   & 3   & 0.20   & 0.006 & 0.5    & 2          & 0.03       & 0.3            \\
l)    & 1.04 & 9.8  & 8.7   & 10  & 0.02   & 0.023 & 1.8    & 2          & 0.02       & 0.9            \\
m)    & 1.47 & 9.8  & 8.9   & 10  & 0.10   & 0.016 & 1.3    & 3          & 0.03       & 0.8		\\  
\end{tabular}
\end{table*}

We have searched through group catalogues of Illustris-1 simulation for possible compact dwarf candidates in the vicinity of the most massive galaxies of clusters of galaxies. We searched for candidates within the distance of 106 kpc (75 kpc/h) from the most massive cluster galaxies which had half-mass radius lower than 1.4 kpc (1 kpc/h). From that search we have identified 22 candidate galaxies which populate the area between UCDs and dEs in galaxy mass-size diagram (Fig \ref{fig:figure1}). To better match the observational results we have used the 2D projected half-mass radius, which was calculated by averaging 1000 projections on a randomly positioned plane.

Unfortunately for almost all of the compact dwarf candidates (from both identified populations) there were no merger tree or progenitor information, therefore we have reconstructed them manually.

For each one of the compact dwarf candidates, type of the particles, particle IDs and phase space information (position, velocities) where extracted from the snapshot data. 

Phase space information and time between snapshots have been used to identify most probable location of the candidate in previous snapshot, after which in a wide area (up to 400 kpc/h radius) around that location candidates for progenitor were extracted from group catalog of previous snapshot in question. Particle ID's for each particle type (stars, gas, dark matter) were cross-referenced with particles from all the progenitor candidates, in order to pinpoint the exact object which is the actual progenitor. The whole procedure was repeated for each following snapshot, for as far long as it was possible towards the beginning of the simulation.

Group catalog data from linked subhalos at each snapshot gave us an insight into the evolution of the subhalos. Out of 22 candidates, 3 were at SUBFIND detection limit (insufficient number of particles) and did not have any structures identified in any of the previous snapshots so they were ruled out from further analysis. For the rest of them there was sufficient data to reconstruct the mass history, size evolution, star formaton history and evolution of their components (dark matter, gas, stars).

Considering that one of the populations (in-situ cluster formed) was created out of the clump of gas inside the cluster (and without dark matter associated with it) we have checked the formation time of stellar particles to confirm that there is no older star particles present, that is - all of stellar particles of that population have been created by star formation process out of the initial gas cloud inside the cluster.


Additionally, for the in-situ population's earliest detected progenitor (gas cloud) we have extracted the following information: the  neutral hydrogen abundance or more precisely fraction of the gas that is in the neutral hydrogen per cell - $HI$, internal energy per unit of mass $u$,  the star-formation rate $SFR$  and the metallicity $Z$. We have expressed metallicity in solar units as: $Z / Z_{0}$, where $Z_{0} = 0.0127$. We have also calculated the temperature $T$ as:
\begin{equation}
  T = \dfrac{2 U \mu m_{H}}{3 k_{B}}
\end{equation}
where $U$ is the internal energy, $\mu$ is the mean molecular weight of the gas, $m_{H}$ is the mass of hydrogen atom, and $k_{B}$ is the Boltzmann constant.

\begin{figure*}
	\includegraphics[width=1.0\textwidth]{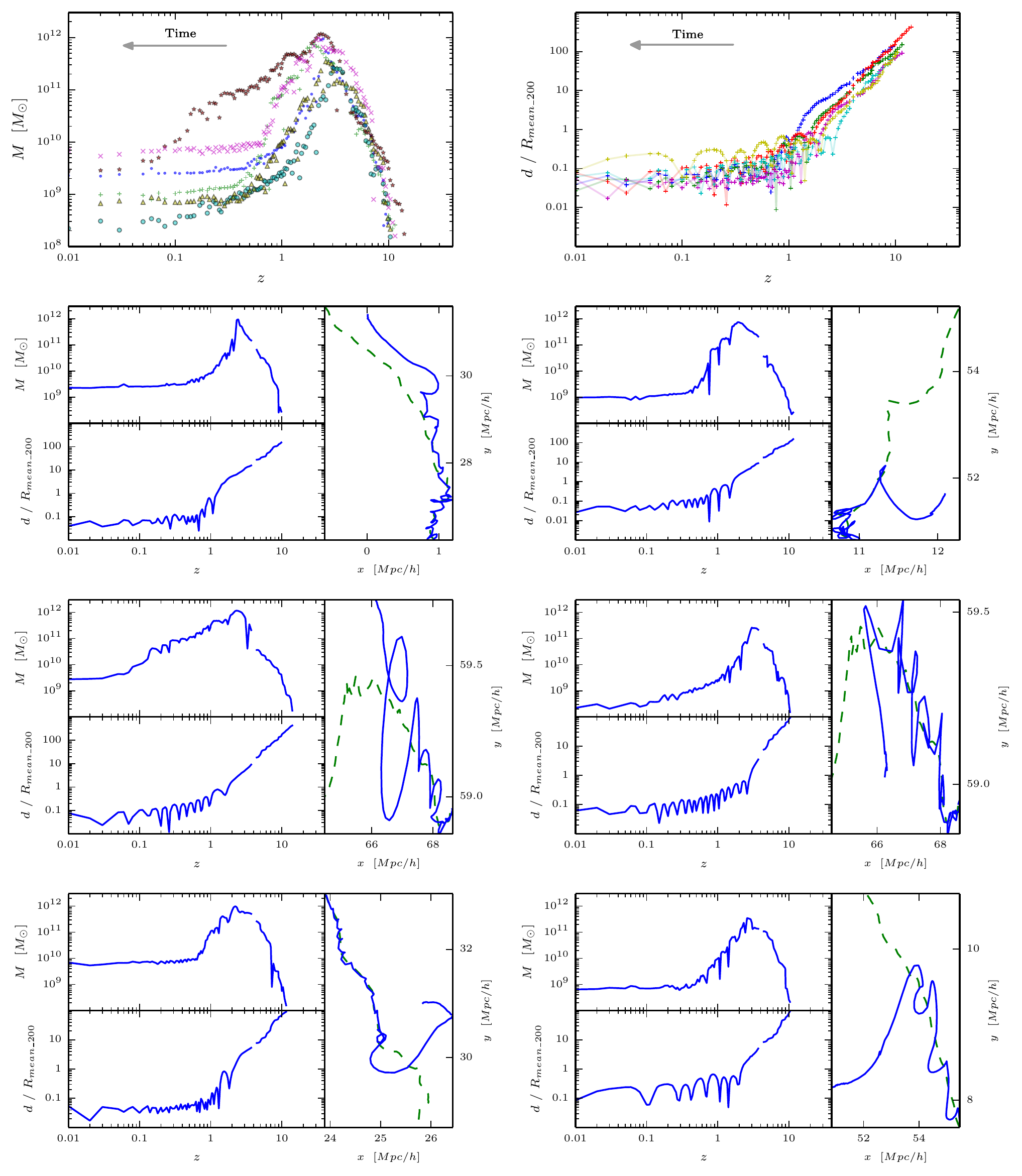}
      \caption{External formation population. Evolution of mass and distance of compact dwarfs from the center of the cluster, relative to the cluster's $R_{mean\_200}$ in a same snapshot. Top row represents the total mass and distance evolution, combined for all candidates. Candidates start loosing mass as they enter the cluster at redshift $z \sim 2$. By redshift $z \sim 1$ they are already deep inside the cluster as compact dwarf galaxies.  Each of the compact dwarf candidates is featured on its own set of mass and distance evolution plots for clear overview. Additional orbit in x-y plane is outlined for easier visualization, where blue line represents motion of dwarf candidate and green dashed line motion of most massive galaxy inside of cluster of galaxies.}
    \label{fig:figure2}
\end{figure*}

\section{Results}

Our interest was to look for the compact dwarf galaxy candidates in the vicinity (inside of 106 kpc) of the most massive galaxies in clusters at $z=0$ in Illustris-1 simulation. 

We have found 19 candidates, which are presented in Fig. \ref{fig:figure1}, where we see their position amongst the rest of the compact dwarf galaxy family on a galaxy mass-radius diagram. Data from the Fig. \ref{fig:figure1} are from \citet{Norris2014} in which compact galaxies are given with half-light radius, where our candidate data is given with 2D projected half-mass radius (averaged over 1000 random projections) from the Illustris-1 simulation.

From the 19 candidates for which the mass history was reconstructed, two populations with different formation history for the compact dwarfs candidates emerged. Parameters for each population are given in Table \ref{tab:table}. Within each population compact dwarf candidates had a very similar history.

In Fig. \ref{fig:figure2} and Fig. \ref{fig:figure4} we have identified a population whose formation stems from the tidal stripping of the Milky Way mass galaxies that have been created externally in reference to the final cluster of galaxies (where we are referencing that the peak of the total virial mass of the galaxies in this population is similar to the estimated total virial mass of the Milky Way $\sim 10^{12} M_{\odot}$ \citep{Kafle2014, McMillan2017}). As can be seen from the plots, formation at higher redshifts follows the standard bottom-up $\Lambda$CDM paradigm. From Fig. \ref{fig:figure4} we can see that objects start their evolution with the expected 20\%/80\% ratio of gas/dark matter, after which there is a build up of mass which proceeds with an active star formation rate. Gas is depleted while stars are formed. However, the expected evolution of these  Milky Way mass galaxies is stopped, and they start loosing their mass as they enter clusters at $z \sim 2$. Most of the mass loss due to the tidal stripping occurs by redshift $z \sim 1$ during interactions with the environment while galaxies sink deeper into the clusters. Due to these interactions galaxies are left without gas and dark matter, where only the compact stellar remnant remains. This coincides with the galaxy reaching the central parts of the cluster where it becomes a close satellite of the most massive cluster galaxy. On the right hand side of each of the subplots of Fig \ref{fig:figure2} a preview of orbits were given for visualization purposes (blue line represents the trajectory of the compact dwarf candidate, while the dashed green line represents the trajectory of the central massive object.)

In Fig. \ref{fig:figure3} and Fig. \ref{fig:figure5} we have identified a compact dwarf candidates for which initial formation happens in-situ in reference to the final cluster. As can be seen from the plots, a dwarf mass galaxy is formed inside the cluster and at lower redshifts ($z \sim 1$). Fig. \ref{fig:figure5} shows that compact dwarf candidates of this population are formed from a gas clouds which are situated inside clusters with no dark matter component. Those clouds of gas undergo short star formation phase which depletes the gas and creates the remaining stellar components. At the end of the process only compact star cores are present. Since they have less time to interact and are already close to the center of their parent clusters, their loss of mass is much smaller than that of the dwarf galaxies in the first population. This population of compact dwarf candidates also end up as a close satellites of massive central objects of the parent clusters.

We address the formation of the in-situ population by analyzing the state of the earliest dwarf progenitors. Parameters for each of the progenitors are extracted and calculated (for solar metallicity and temperature), and presented in Table \ref{tab:table2}. Additionally HI abundance, metallicity in solar units, temperature and stellar content maps are given in Fig. \ref{fig:figure6}, \ref{fig:figure7} and \ref{fig:figure8}.

Results from the Table \ref{tab:table2} and Fig. \ref{fig:figure6}, \ref{fig:figure7} and \ref{fig:figure8} show that all gas clouds are composed out of cold ($< 10^{5}$ K), dense gas at various phases of star formation. Temperature and mass ranges of the gas clouds are consistent with the cold-mode accretion in various phases \citep{Keres2005, Norris2014} as a source of formation.

Three gas clouds are in the earliest phases of star formation (f), j) and k)), as they are just starting to form stellar content. As such they are distinguished from other clouds by higher HI abundances, slightly lower temperatures and lower SFR. Metallicities of all gas clouds are slightly higher but still consistent with results from earlier works \citep{Genel2016, Kacprzak2016} at the specific redshift, amount of cold gas, and star formation phase of the gas clouds. Those gas clouds where star formation has started recently will have the same destiny as the ones that are already undergoing significant SFR. That is, they will experience depletion of the gas by star formation, enrichment the surrounding gas and an increase in metallicity (as can be concluded from Fig. \ref{fig:figure6}, \ref{fig:figure7} and \ref{fig:figure8}).

Ultimately both external and in-situ populations form similar compact dwarf candidates found at $z=0$.

\section{Discussion}

\begin{figure*}[p]
	\includegraphics[width=1.0\textwidth]{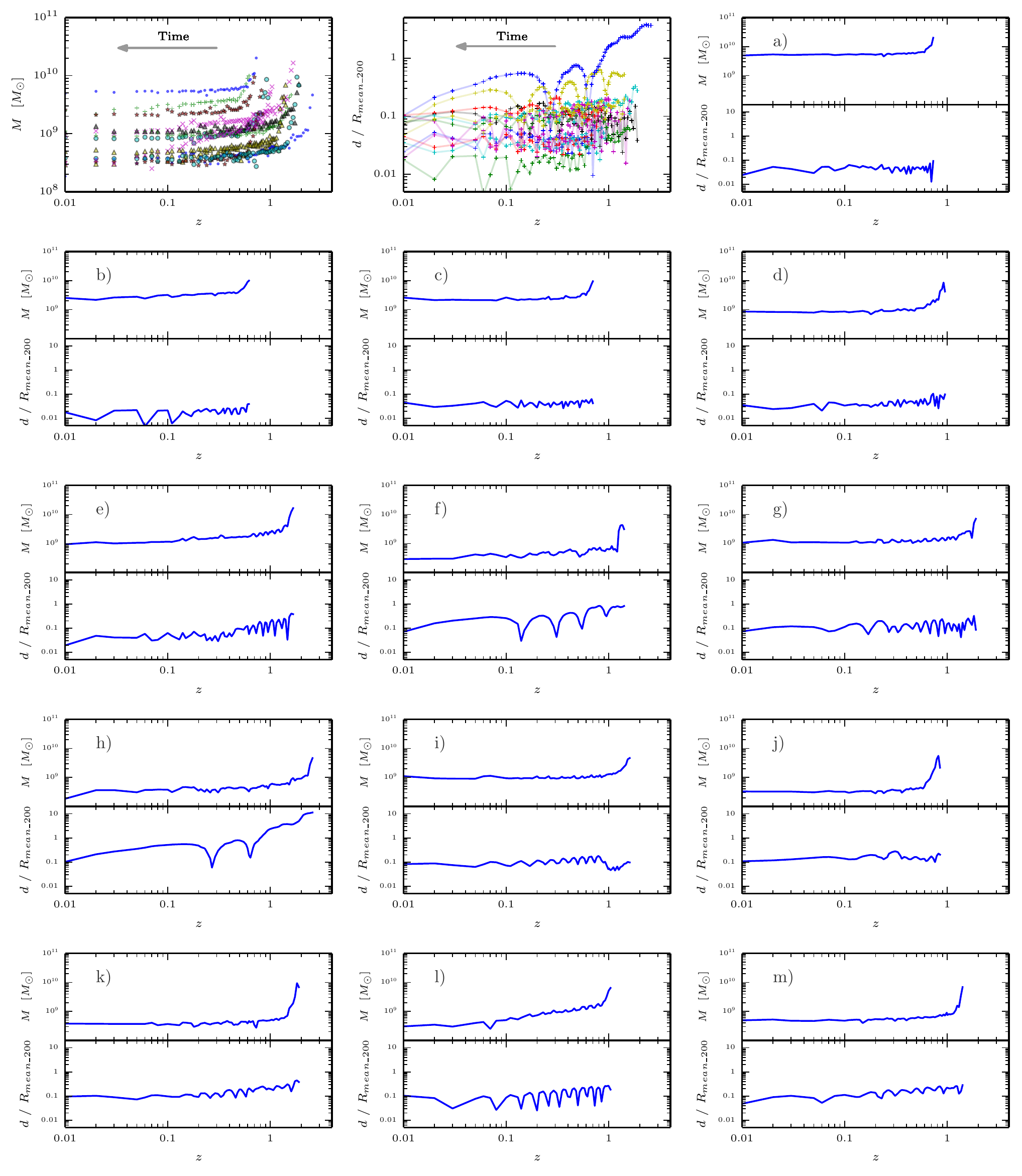}
    \caption{In-situ cluster formed population. Evolution of mass and distance of compact dwarfs from the center of the cluster, relative to the cluster's $R_{mean\_200}$ in a same snapshot. Similar as in Fig. \ref{fig:figure2}, first two plots of top row represent the total mass and distance evolution, combined for all candidates. Here we can see that galaxies in this population form within clusters and have less dramatic mass evolution than external population presented in Fig. \ref{fig:figure2}. Each of the compact dwarf candidates is featured on its own set of mass and distance evolution plots for clear overview. Orbit outline is omitted here because of additional area it would necessitate.}
    \label{fig:figure3}
\end{figure*}

\begin{figure}
	\includegraphics[width=\columnwidth, height=20cm]{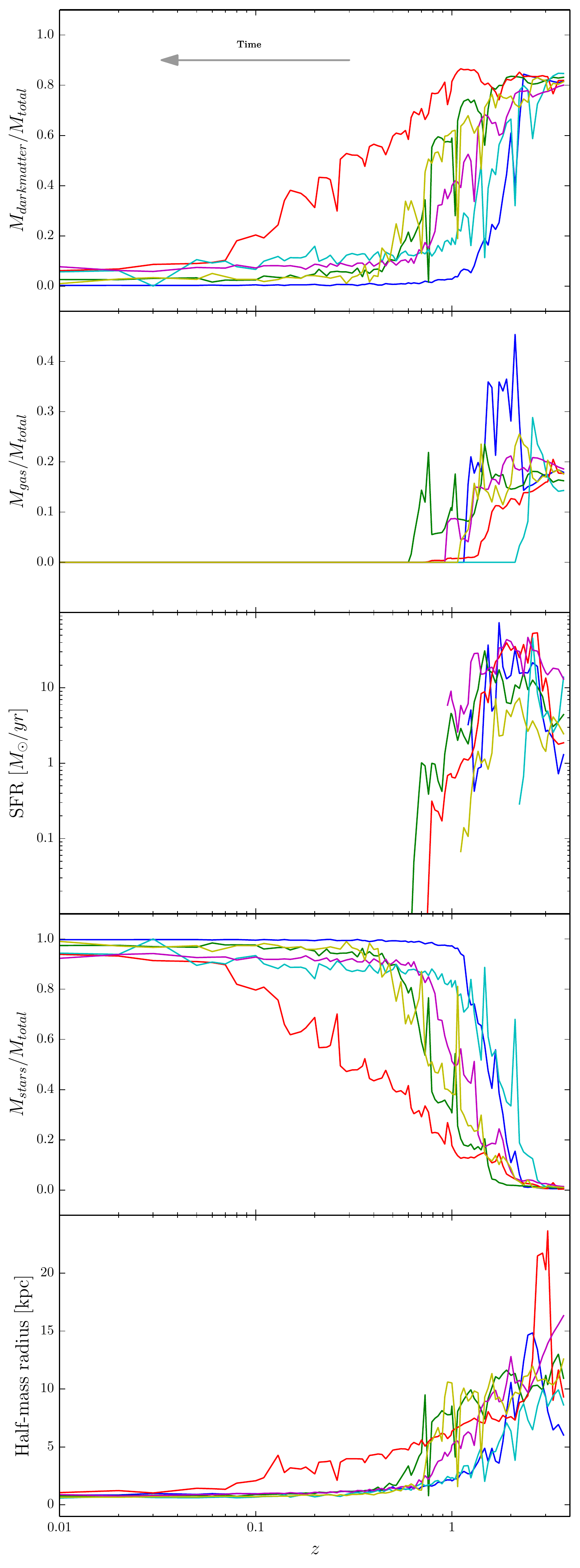}
      \caption{Mass and size evolution of external formation population. From top to bottom: evolution of dark matter mass fraction, evolution of gas mass fraction, star formation rate within twice the stellar half-mass radius, evolution of star mass fraction and evolution of size given as \text{red}{stellar} half-mass radius. It is clearly seen that after standard $\Lambda CDM$ bottom-up paradigm Milky Way size and mass like galaxies are formed after which they are stripped (upon entry into clusters) to compact stellar remnants. }
    \label{fig:figure4}
\end{figure}

Two groups of candidate compact dwarf galaxies at $z=0$ from Illustris-1 simulation are identified by their different formation histories.

External origin population are tidally stripped Milky Way mass galaxies. They were formed outside of the clusters in a manner that is characteristic for the $\Lambda CDM$ bottom-up paradigm. They are created at very low redshifts as a mixture of gas and dark matter (20\%/80\%). Accumulation of mass is followed by star formation that depletes the gas and creates stars. But after the in-fall into the cluster their mass evolution suggest that these Milky Way mass objects are exposed to extreme tidal stripping. In the meantime, change in half-mass radius points to transformation of a Milky Way size object to a compact dwarf candidate. Both gas and dark matter are completely stripped away (or in the case of the gas - depleted as well) leaving only the compact stellar remnant behind. All of the candidates ultimately become close satellites of the most massive galaxy in the cluster, which can be clearly seen on the orbital outline.

In-situ origin population are compact dwarf candidates which form within the clusters. Their formation masses are characteristic for dwarf galaxies.

Parameters extracted from the earliest detected progenitor - a gas cloud undergoing star formation - are consistent with the cold-mode accretion picture  \citep{Keres2005}. Considering how deep inside clusters progenitor gas clouds were detected for the first time, this points either to the possibility that some of the gas cooled down near cluster center and was blown away by AGN activity \citep{Nipoti2004} or it points towards an external origin (in reference to the cluster) of the gas (driven by filaments in order to reach deep inside the cluster). Unfortunately time resolution of the simulation prevents us from conclusively determining the true nature of the gas origin. Gas probably started in the intergalactic matter in diffuse phase and moved directly to the dense galactic phase without heating more than $\sim 10^{5}$ K \citep{Keres2005}. Within the cold gas there is a SFR process in all progenitor gas clouds, but apparently in different phases. Some are well under way, while other clouds are detected just before star formation process is starting (most notably f) and k) labeled gas clouds) as is illustrated from  Fig. \ref{fig:figure6}, \ref{fig:figure7} and \ref{fig:figure8}.

Unlike the external population, in-situ population candidates have their growth suppressed from the start by the environment inside the cluster. They are created in more recent times and from the start they are devoid of dark matter (although it should be noted that they are immersed within dark matter halo of parent cluster). Initially they are merely a compact cloud of gas which goes through a short star formation phase while creating stars and depleting all of the gas. Compact stellar objects are created after the star formation phase. They too become satellites of the most massive galaxies, but experience less dramatic mass loss than the external population.

Looking at the Fig. \ref{fig:figure3} one may notice that h) designated compact dwarf candidate is created outside of the cluster unlike other members of that population. But all other parameters of this interesting compact object are consistent with the in-situ population. Further analysis reveals that this object was created within a less massive group in which it interacted with its most massive object. Eventually both objects (massive galaxy and its compact companion) have ended up interacting, as other members of the in-situ population, with the most massive galaxy in final parent cluster. As such, its history is not straightforward but in essence very similar in nature to the rest of the population.

Both populations descent to form close satellites of the most massive galaxies of the clusters, which is somewhat expected considering that the initial filter was to search for the compact dwarf candidates at those location. However, no similar candidates on flyby trajectories are found, which, at least for external population, hints that long term tidal stripping inside cluster is essential for creating compact dwarf candidates which ultimately end up as satellites around the most massive galaxies.

\begin{figure}
	\includegraphics[width=\columnwidth, height=20cm]{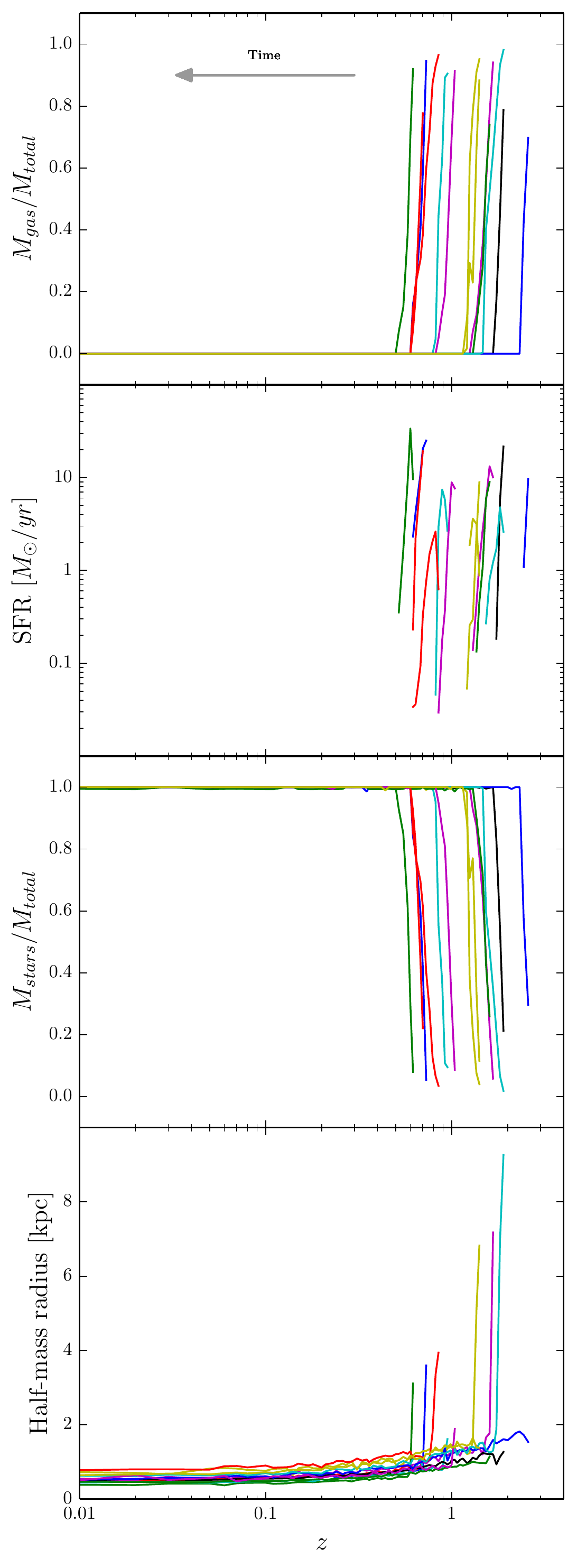}
      \caption{Mass and size evolution of in-situ cluster formed population. From top to bottom: evolution of gas mass fraction, star formation rate within twice the stellar half-mass radius, evolution of star mass fraction and evolution of size given as stellar half-mass radius. These populations are formed much later in the simulation within cluster as a compact gas cloud which goes through short star formation phase after which gas is depleted and compact stellar component remains. It should be noted that they do not have dark matter component associated with them.}
    \label{fig:figure5}
\end{figure}

We have searched for compact dwarf candidates by locating them on a galaxy mass-size diagram. Note that half-light radius of galaxies retrieved from observations is compared against half-mass radius from simulation. Connection between the two is not straightforward (unless when M/L ratio is constant) and depends both on compactness of the object and from their mass. For compact lower mass objects half-light radius tends to be lower than half-mass radius (e.g. \citet{Wellons2015}) which is referenced with arrows in Fig. \ref{fig:figure1}. Thus, although our compact dwarf candidates are higher on the plot, they would be expected to move further down the regime of cEs and higher mass UCDs, given that the half-light radius is expected to be lower in this case.

We have found 19 compact dwarf candidates from 14 clusters. This result is consistent with the preliminary results by \citet{Zhang2017}. They observed 17 clusters of galaxies and found ~ 45 compact dwarfs inside 300 Mpc and ~ 10 inside 50 kpc from the brightest galaxies in the clusters.

It should be noted as a future research opportunity that second most massive cluster in Illustris-1 did not yield a single candidate for compact dwarfs at $z=0$ suggesting that cluster evolution itself could play a role in forming compact dwarfs in the vicinity of its most massive galaxies. 

In total, around 30\% of candidates in the vicinity of most massive galaxies in clusters have been formed by tidal stripping of Milky Way mass galaxies infalling into the cluster, while the rest (around 70\%) are in-situ cluster formed from dwarf like progenitors. This is the most important result of our paper considering that observations can not easily distinguish various populations of compact dwarfs.

It should be also noted that considering that gravitational softening for Illustris-1 for baryons at $z=0$ is 710 pc, and the hydrodynamical gas cell resolution is 48 pc, it is probably not possible to retrieve more compact objects than the ones found. Thus it is reasonable to assume that these represent top of the population of compact dwarf galaxies near most massive galaxies in clusters. One would expect more of compact objects to be found when future higher resolution full physics cosmological simulations with similar adequate analysis are made available.

\section*{Acknowledgements}

We would like to thank the anonymous reviewer for his/hers careful reading of our manuscript and suggestions that helped to improve the paper. We would also like to thank Jovana Petrovi\v{c}, Darko Donevski and Milan \'{C}irkovi\'{c} for useful discussions.
This work was supported by the Ministry of Education, Science and Technological Development of the
Republic of Serbia through project no. 176021, ``Visible and Invisible Matter in Nearby Galaxies:
Theory and Observations.''

\begin{figure*}
	\includegraphics[width=1.0\textwidth]{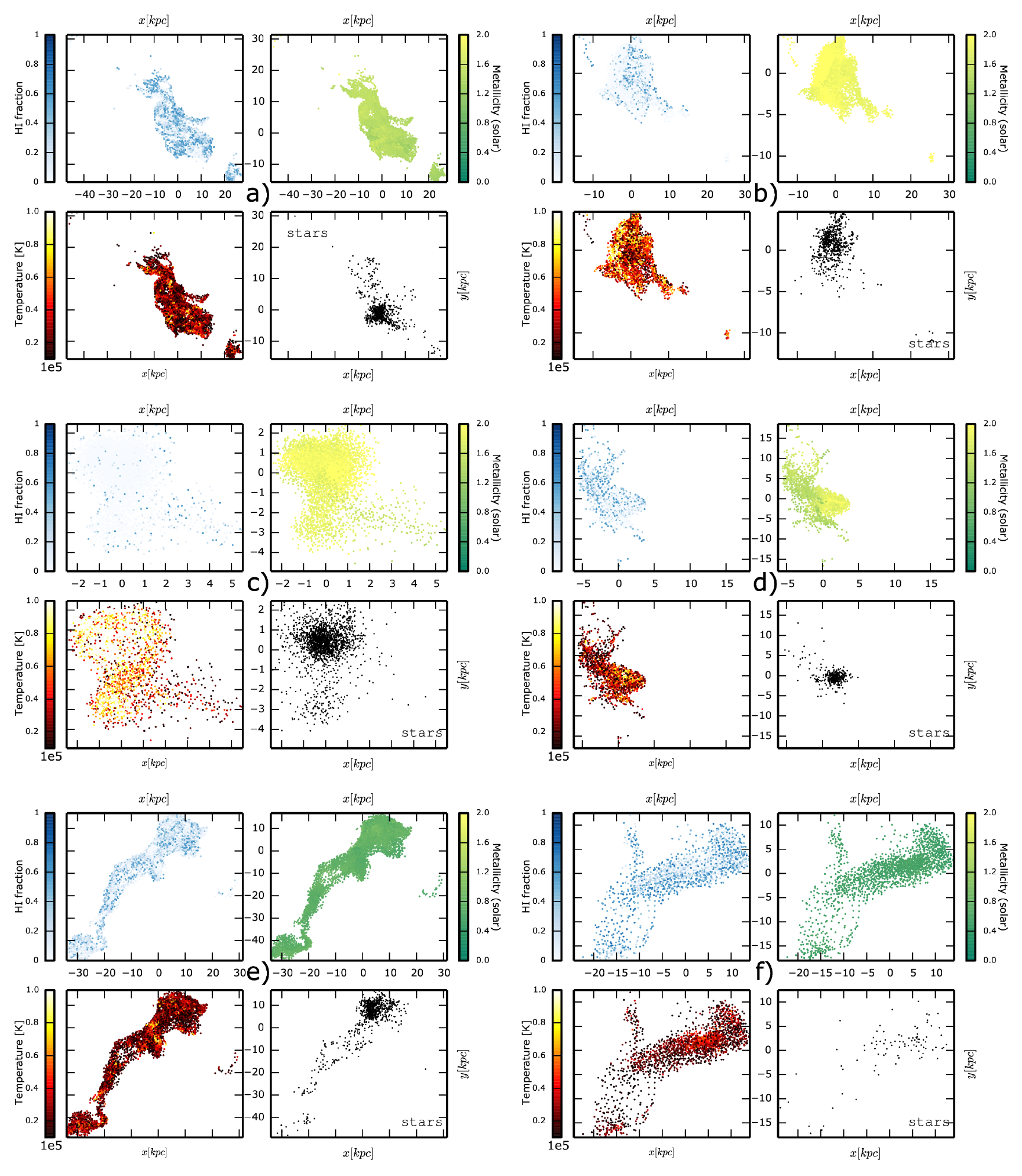}
      \caption{In-situ formation population progenitors. Spatial distribution for HI abundance of the gas (top left), metallicity in solar units of the gas (top right), temperature of the gas (lower left) and stellar content (lower right). All the plots are for earliest detected gas clouds. Labels correspond with labels in Tables \ref{tab:table} and \ref{tab:table2}.}
    \label{fig:figure6}
\end{figure*}

\begin{figure*}
	\includegraphics[width=1.0\textwidth]{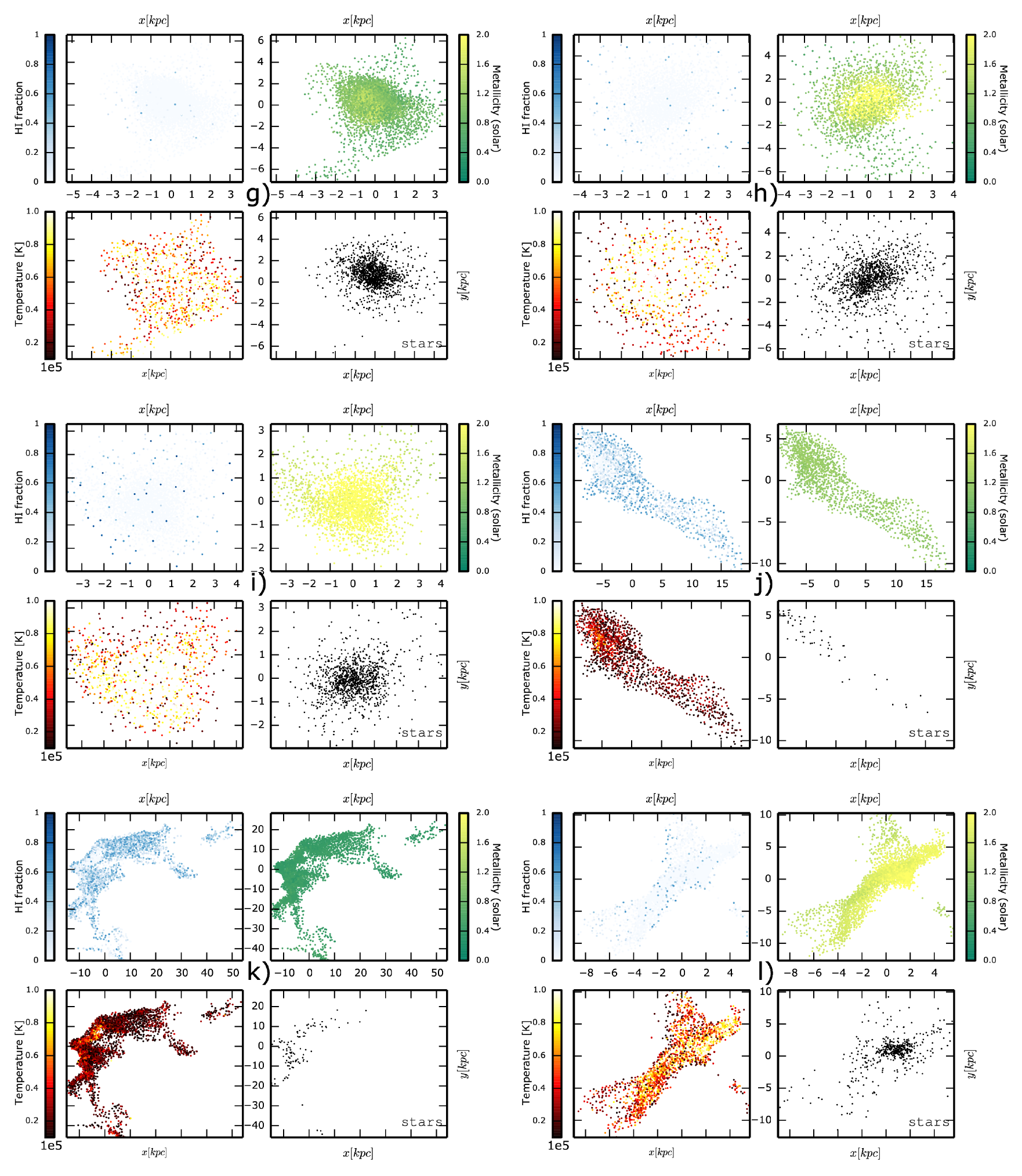}
      \caption{In-situ formation population progenitors (continued from Fig. \ref{fig:figure6}). Spatial distribution for HI abundance of the gas (top left), metallicity in solar units of the gas (top right), temperature of the gas (lower left) and stellar content (lower right). All the plots are for earliest detected gas clouds. Labels correspond with labels in Tables \ref{tab:table} and \ref{tab:table2}.}
    \label{fig:figure7}
\end{figure*}

\begin{figure*}
	\includegraphics[width=1.0\textwidth]{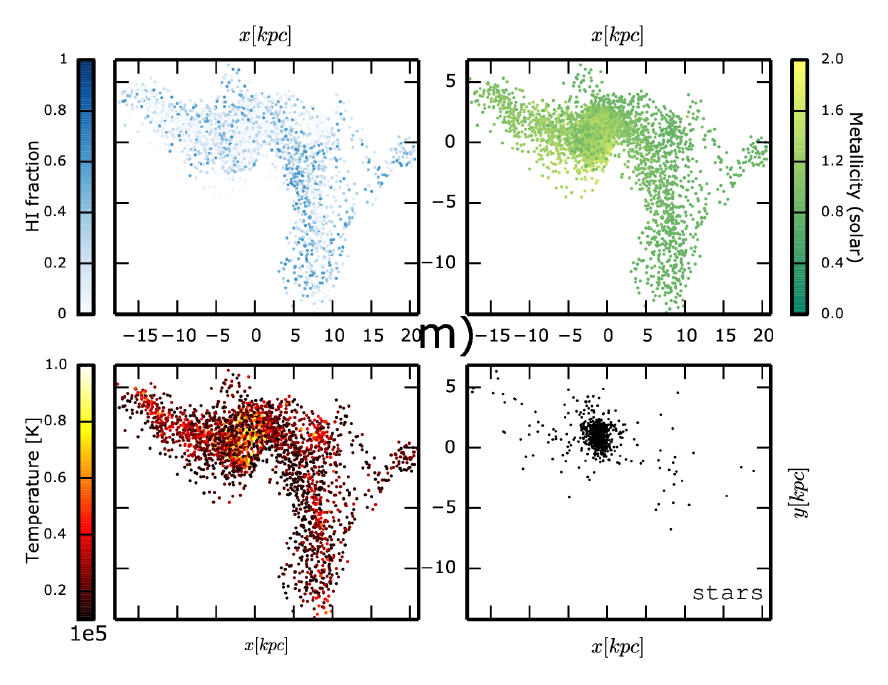}
      \caption{In-situ formation population progenitors (continued from Fig. \ref{fig:figure7}). Spatial distribution for HI abundance of the gas (top left), metallicity in solar units of the gas (top right), temperature of the gas (lower left) and stellar content (lower right). All the plots are for earliest detected gas clouds. Labels correspond with labels in Tables \ref{tab:table} and \ref{tab:table2}.}
    \label{fig:figure8}
\end{figure*}



\bsp	

\label{lastpage}

\begin{thebibliography}{99}

\bibitem[\protect\citeauthoryear{Bekki et al.}{2001}]{Bekki2001}
Bekki K., Couch W. J., Drinkwater M. J., Gregg M. D., 2001, ApJ, 557, L39

\bibitem[\protect\citeauthoryear{Bekki et al.}{2003}]{Bekki2003}
Bekki K., Couch W. J., Drinkwater M. J., Shioya Y., 2003, MNRAS, 344, 399

\bibitem[\protect\citeauthoryear{Chilingarian \& Mamon}{2008}]{chilingarian2008}
Chilingarian I. V., Mamon G. A., 2008, MNRAS, 385, L83

\bibitem[\protect\citeauthoryear{Chilingarian et al.}{2007}]{Chilingarian2007}
Chilingarian I., Cayatte V., Chemin L., Durret F., Lagan\'{a} T. F., Adami C., Slezak E., 2007, A\&A, 466, L21

\bibitem[\protect\citeauthoryear{Drinkwater et al.}{2000}]{Drinkwater2000}
Drinkwater M. J., Jones J. B., Gregg M. D., Phillipps S., 2000, PASA, 17, 227 

\bibitem[\protect\citeauthoryear{Drinkwater et al.}{2003}]{Drinkwater2003}
Drinkwater M. J., Gregg M. D., Hilker M., Bekki K., Couch W. J., Ferguson H. C., Jones J. B., Phillipps S., 2003, Nature, 423, 519

\bibitem[\protect\citeauthoryear{Fellhauer \& Kroupa}{2002}]{Fellhauer2002}
Fellhauer M., Kroupa P., 2002, Ap\&SS, 281, 355

\bibitem[\protect\citeauthoryear{Genel}{2016}]{Genel2016}
Genel S., 2016, ApJ, 822, 107

\bibitem[\protect\citeauthoryear{Hilker et al.}{1999}]{Hilker1999}
Hilker M., Infante L., Richtler T., 1999, A\&AS, 138, 55

\bibitem[\protect\citeauthoryear{Jones et al.}{2006}]{Jones2006}
Jones J. B. et al., 2006, AJ, 131, 312

\bibitem[\protect\citeauthoryear{Kacprzak et al.}{2016}]{Kacprzak2016}
Kacprzak G. G. et al., 2016, ApJ, 826, L11

\bibitem[\protect\citeauthoryear{Kafle et al.}{2014}]{Kafle2014}
Kafle P. R., Sharma S., Lewis G. F., Bland-Hawthorn J., 2014, ApJ, 794, 17

\bibitem[\protect\citeauthoryear{Keres et al.}{2005}]{Keres2005}
Kere\v{s} D., Katz N., Weinberg D.H., Dav\'{e} R., 2005, MNRAS, 363, 2

\bibitem[\protect\citeauthoryear{McMillan}{2017}]{McMillan2017}
McMillan P. J., 2017, MNRAS, 465, 76

\bibitem[\protect\citeauthoryear{Mieske, Hilker \& Infante}{2002}]{Mieske2002}
Mieske S., Hilker M., Infante L., 2002, A\&A, 383, 823

\bibitem[\protect\citeauthoryear{Mieske et al.}{2004}]{Mieske2004}
Mieske S. et al., AJ, 2004, 128, 1529

\bibitem[\protect\citeauthoryear{Mieske et al.}{2005}]{Mieske2005}
Mieske S., Infante L., Hilker M., Hertling G., Blakeslee J. P., Ben\'{\i}tez N., Ford H., Zekser K., 2005, A\&A, 430, L25

\bibitem[\protect\citeauthoryear{Misgeld et al.}{2011}]{Misgeld2011}
Misgeld I., Mieske S., Hilker M., Richtler T., Georgiev I. Y., Schuberth Y., 2011, A\&A, 531, 16

\bibitem[\protect\citeauthoryear{Nelson et al.}{2015}]{Nelson2015}
Nelson D. et al., 2015, A\&C, 13, 12

\bibitem[\protect\citeauthoryear{Nipoti \& Binney}{2004}]{Nipoti2004}
Nipoti C., Binney J., 2004, MNRAS, 349, 1509

\bibitem[\protect\citeauthoryear{Norris et al.}{2014}]{Norris2014}
Norris M. A., 2014, MNRAS, 443, 1151

\bibitem[\protect\citeauthoryear{Penny et al.}{2012}]{Penny2012}
Penny S. J., Forbes D. A., Conselice C. J., 2012, MNRAS, 422, 885

\bibitem[\protect\citeauthoryear{Penny et al.}{2014}]{Penny2014}
Penny S. J., Forbes D. A., Strader J., Usher C., Brodie J. P., Romanowsky A.J., 2014, MNRAS, 439, 3808

\bibitem[\protect\citeauthoryear{Price et al.}{2009}]{Price2009}
Price J. et al., 2009, MNRAS, 397, 1816

\bibitem[\protect\citeauthoryear{Rodriguez-Gomez et al.}{2015}]{Rodriguez2015}
Rodriguez-Gomez V. et al., 2015, MNRAS, 449, 49

\bibitem[\protect\citeauthoryear{Springel et al.}{2001}]{Springel2001}
Springel V., White S. D. M., Tormen G., Kauffmann G., 2001, MNRAS, 328, 726

\bibitem[\protect\citeauthoryear{Vogelsberger et al.}{2014}]{Vogelsberger2014}
Vogelsberger M. et al., 2014, MNRAS, 444, 1518

\bibitem[\protect\citeauthoryear{Wellons et al.}{2015}]{Wellons2015}
Wellons et al., 2015, MNRAS, 449, 361

\bibitem[\protect\citeauthoryear{Wellons et al.}{2016}]{Wellons2016}
Wellons S. et al., 2016, MNRAS, 456, 1030

\bibitem[\protect\citeauthoryear{Zhang \& Bell}{2017}]{Zhang2017}
Zhang Y., Bell E. F., 2017, ApJL, 835, 6


\end{thebibliography}
\end{document}